# Formation process of skyrmion lattice domain boundaries: The role of grain boundaries

H. Nakajima[1], A. Kotani[1], M. Mochizuki[2], K. Harada[1, 3], and S. Mori[1]

[1]*Department of Materials Science, Osaka Prefecture University, Sakai, Osaka 599-8531, Japan*

[2]*Department of Applied Physics, Waseda University, Shinjuku, Tokyo 169-8555, Japan*

[3]*Center for Emergent Matter Science, the Institute of Physical and Chemical Research (RIKEN), Hatoyama, Saitama 350-0395, Japan*

We report on the formation process of skyrmion lattice (SkL) domain boundaries in FeGe using Lorentz transmission electron microscopy and small-angle electron diffraction. We observed that grain boundaries and edges play an important role in the formation of SkL domain boundaries; The SkL domain boundary is stabilized at the intersection of two grains. A micromagnetic simulation using the Landau−Lifshitz−Gilbert equation revealed that the SkL domains separated by a boundary represent the lowest energy configuration. Conversely, in a wide area, SkL domain boundaries were not formed and SkL domains with different orientations rotated to form a single SkL domain.

Magnetic skyrmions are topologically stable vortex-like objects that can be stabilized via the Dzyaloshinskii–Moriya (DM) interaction in chiral magnets[1]. They have been observed in *B*20-type cubic magnets with a chiral space group $P2_13$, such as MnSi, $Fe_{1-x}Co_xSi$, and FeGe, by means of small-angle neutron scattering and Lorentz transmission electron microscopy (Lorentz microscopy)[2–4]. Skyrmions exhibit a number of quantum phenomena such as the topological Hall effect and magnetoelectric coupling[5–7]. Furthermore, recent discoveries of manipulation of skyrmions by low electric currents or fields indicate that skyrmions are suitable for the development of magnetic memory devices with high density and low power consumption[8–10].

Skyrmions crystallize into a triangular lattice with hexagonal symmetry called the skyrmion lattice (SkL). Magnetic structures usually behave as continuous spin textures. However, skyrmions exhibit a particle-like nature that is suitable for data storage and spintronic devices[11]. In fact, single skyrmion particles have been observed in a certain range of magnetic fields (40 ~ 80 mT)[12]. Similar to solid atoms in a crystal lattice, SkLs form boundaries between SkL domains. In relation to the particle-like nature of skyrmions, SkL domain boundaries are one of the most important characteristics because they can give direct evidence of the particle-like nature[13, 14]. SkL domain boundaries have been observed via several methods such as small-angle resonant x-ray diffraction and Lorentz microscopy. An x-ray diffraction study indicated that SkL domains with different orientations can be formed by tilting the magnetic field[15]. Another study based on Lorentz microscopy reported that SkL domain boundaries contain a pair of skyrmions with seven and five neighbors[16]. Furthermore, a microscopy study based on differential phase contrast revealed that skyrmions are elongated at SkL domain boundaries[17]; this observation highlights not only the particle-like character but also the topological properties of skyrmions, which undergo continuous deformations without changing their topological number[14].

These properties of SkL domain boundaries are also important to differentiate skyrmions from other vortices and to understand the particle-like nature of skyrmions. Several other systems are known to form triangular lattices and boundaries at their domains such as bubble rafts, colloidal crystals, superconducting vortices (fluxons), and crystal lattices[18–24]. However, compared to these systems, SkL domain boundaries are peculiar as they fluctuate over time and can be moved by the magnon currents[13, 16]. Another difference between skyrmions and fluxons, which are both magnetic objects, lies in the structures of domain boundaries. Fluxons also form triangular lattices as a ground state[20]. In addition, various domain boundary structures of the vortex lattices have been observed by Lorentz microscopy[21, 22]. However, these lattice domains show linear chains of fluxons by tilting magnetic fields or random distribution of fluxons by defects, while topological skyrmions present arrangements of vortices with five and seven neighbors at the domain boundary[16].

Although several unique features of SkL domain boundaries have been reported[13–17], their formation process has never been monitored. However, it is important to understand the formation processes and to reveal the origins of SkL domain boundaries because they can clarify the difference between skyrmions (the particle-like objects) and helical domains (continuous spin textures). In such an observation, we can also reveal how the continuous spin character changes into the particle-like nature with an increasing magnetic field.

In this paper, we investigated the formation process of SkL domain boundaries in areas confined by grain boundaries using Lorentz microscopy. In the area confined by grain boundaries, SkL domains are forced to align along the grain boundaries, thus resulting in the formation of a SkL domain boundary. Furthermore, the SkL domain is formed in an area where no helical domains are present, indicating that the particle-like nature is acquired across the phase transition from the helical to skyrmion states. In another case where skyrmions are formed in a wide area, SkL domains with different orientations rotate to form a single triangular SkL domain instead of creating a boundary when they collide during the application of an increasing magnetic field.

Polycrystals of FeGe were synthesized via arc melting and high-pressure treatments as described in Ref. 4. The temperature dependence of the magnetization was measured using a vibrating sample magnetometer. Thin specimens for Lorentz microscopy observations were prepared via Ar$^+$ ion milling after a mechanical polishing procedure. We performed Lorentz microscopy based on the Fresnel method and small-angle electron diffraction (SmAED) using a transmission electron microscope (TEM) JEM-2100F operated at 200 kV ($\lambda$ = 2.51 pm). In the Fresnel method, real space distributions of magnetic domain structures are visualized as bright and dark contrast due to the Lorentz deflection. Small-angle electron diffraction is an effective method for the quantitative analysis of magnetic domain structures[25–27]. The diffraction pattern contains spots from the phase grating of the periodic magnetic structure as well as the magnetic deflection due to the Lorentz force[28,29]. SmAED has several advantages over small-angle neutron and x-ray diffraction techniques because specific nanoscale or microscale areas can be selected for diffraction owing to the electromagnetic lenses and apertures. In addition, a real-space image can be obtained in the same area as the diffraction pattern. This SmAED technique has been used to study 180° domains in ferromagnets, rotating magnetization in a cobalt particle, and magnetization processes in a Co/Cu multilayer film[28–31]. In this study, SmAED in external magnetic fields was performed using an electron optical system with a long camera length[32,33]. A magnetic field was applied perpendicular to the thin specimen using the objective lens. We corrected the rotations of images due to the Lorentz force originating from the objective lens using the angles measured with a carbon grating after the images were taken. The camera lengths for the electron diffraction were elongated by controlling the currents in the intermediate lenses of the TEM. In the Fresnel images, the specimens were slightly tilted to reduce the bend contour due to the diffraction effect.

The Fresnel method was used to visualize the formation process of the SkL domain boundary in an area surrounded by grain boundaries [Figs. 1(a)–(d)] during the application of an increasing magnetic field. Figure 1(a) shows the helical domain structures at 260 K in the absence of an external magnetic field. The helical domains are surrounded by grain boundaries and the specimen edge. Similar to a previous study[34], the propagation vectors [white arrows] of the helical domains point in various directions owing to the small magnetocrystalline anisotropy. With an increase in magnetic field, the helical structures were pinched from the edges of the helical domains [Figs. 1(b) and 1(c)]. When the hexagonal SkL was formed, the helical structures remained in the center of the grain [see in Fig. 1(c)]. The patches of skyrmions in Fig. 1(c) have no uniform orientation. However, due to the grain boundaries the system is forced to have two orientations in Fig. 1 (d). The SkL domain boundary consists of skyrmions with five and seven neighbors [Fig. 1(e)]. At this boundary, some skyrmions were elongated and typical skyrmions with six neighbors were also distributed between skyrmions with five and seven neighbors. A comparison of Figs. 1(a) and 1(d) indicates that this SkL domain boundary was created in an area where no helical domains were present. Furthermore, each skyrmion moved to be densely arranged within the grain. These two results indicate that the continuous spin texture acquired a particle-like nature across the phase transition from the helical to skyrmion states. We observed that the SkL domain boundary was reproduced in the same region when the specimen was cooled below $T_N$ and then a magnetic field was applied after reheating the specimen to a temperature above $T_N$ = 280 K [see Fig. S1 of the supplementary material]. This SkL domain boundary was not affected by a magnetic field higher than 1 T until the skyrmions changed into a forced ferromagnetic structure; in addition, temperature variations did not produce any movement of the SkL domain boundary. The two SkL domains in Fig. 1(d) are aligned along the grain boundaries, and the SkL domain boundary is located between the two SkL domains. The right side of the SkL domain boundary is located at the intersection of two grains. Therefore, these results indicate that the SkL boundary can be formed in the area confined by grain boundaries.

For a theoretical description of the experimental observations, we perform a micromagnetic simulation using the Landau–Lifshitz–Gilbert (LLG) equation that is represented as follows:

$$\frac{d\boldsymbol{m}_i}{dt} = -\boldsymbol{m}_i \times \boldsymbol{H}_i^{\text{eff}} + \frac{\alpha_G}{m}\boldsymbol{m}_i \times \frac{d\boldsymbol{m}_i}{dt}, \quad (1)$$

where $\alpha_G$ = 0.04 is the Gilbert damping coefficient. The effective field $\boldsymbol{H}_i^{\text{eff}}$ acting on the local magnetization $\boldsymbol{m}_i$ is calculated from the Hamiltonian $\mathcal{H}$ as $\boldsymbol{H}_i^{\text{eff}} = -\partial\mathcal{H}/\partial\boldsymbol{m}_i$. The Hamiltonian for a thin specimen of chiral-lattice ferromagnet is given by the classical Heisenberg model of a square lattice with ferromagnetic-exchange interaction, the DM interaction, and the Zeeman coupling associated with a magnetic field $\boldsymbol{H} = (0, 0, H_z)$ normal to the plane[35] that is given by the following equation:

$$\mathcal{H} = -J\sum_{<i,j>}\boldsymbol{m}_i \cdot \boldsymbol{m}_j - H_z\sum_i m_{iz} - D\sum_i (\boldsymbol{m}_i \times \boldsymbol{m}_{i+\hat{x}} \cdot \hat{\boldsymbol{x}} + \boldsymbol{m}_i \times \boldsymbol{m}_{i+\hat{y}} \cdot \hat{\boldsymbol{y}}), (2)$$

The norm of the magnetization vectors $\boldsymbol{m}_i$ is set to be unity. We use $J$ = 1, $D/J$ = 0.36, and $H_z$ = 0.06 for the calculations.

We first prepare a single domain skyrmion lattice by a Monte Carlo calculation using a square system of 288 × 288 sites with periodic boundary conditions. Then we cut the system into the shape that is shown in Fig. 2(a) and impose an open boundary condition to simulate an area in the thin-plate specimen, which is observed in the experiment, surrounded by two grain boundaries and the specimen edge. Starting with this initial skyrmion configuration, we simulate the spatiotemporal magnetization dynamics of the relaxation process towards the formation of stable domain structures, by numerically solving the LLG equation. The obtained time evolution of the total energy in Fig. 2(b) shows a monotonic decrease over time, thus indicating that the packing of the skyrmions is rearranged to lower the total energy. After sufficient relaxation, we obtain two different skyrmion-lattice domains (A and B) that are separated by a domain boundary as shown in Fig. 2(c). The domains A and B are favored by the vertical and horizontal edges of the system, respectively. We find that the obtained domain structure contains skyrmions with five and seven neighbors at the domain boundary and elongated skyrmions that closely reproduce the experimental observations.

Subsequently, we investigated the formation process of the SkL in a wider area far from grain boundaries using the Fresnel and SmAED methods. Figure 3(a) shows an overfocused Fresnel image at 260 K and zero magnetic field. Helical structures appear similar to those shown in Fig. 1(a). A SmAED pattern from the area labeled as "A" in Fig. 3(a) is displayed in Fig. 3(e). The SmAED pattern shows satellite peaks [red arrows $\theta \sim 3.3 \times 10^{-5}$ rad and white arrows $\theta \sim 6.6 \times 10^{-5}$ rad] around the direct beam; this indicates that the magnetic domains have a helically modulated spin structure. These observed diffraction angles of the spots correspond to a modulated period of ~76 nm that is in agreement with the Fresnel image [Fig. 3(a)]. Note that striped 180° domains show bright and dark contrast in a Fresnel image and satellite peaks in a diffraction pattern similar to helical domains[28, 29, 36]. However, the magnetization measurements reported in the supplementary material showed a clear cusp that is caused by a helical magnetic structure due to the DM interaction. Thus, based on the magnetization measurements in Fig. S2, it can be confirmed that the magnetic domains in Fig. 3(a) are helical magnetic domains. When the magnetic field was further increased, SkL domains emerged from the helical domains [see Fig. 3(b)]. These two domains expanded with an increase in magnetic field [see Fig. 3(c)]. The SmAED patterns for the two regions labeled as "B" and "C" in Fig. 3(c) are shown in Figs. 3(f) and 3(g), respectively. The patterns comprise six spots that result from the SkL (three propagation vectors at relative angles of 120°). The angles of diffraction spots in Fig. 3(f) and 3(g) are 2.0° and 15° from the horizontal line, respectively, and the SkL domains rotated by 13° with respect to each other [see the red and blue arrowheads]. The difference in the orientations of the SkL domains is also indicated by the red dashed lines in Fig. 3(c). The diffraction spots of $3.5 \times 10^{-5}$ rad equal [70 nm]$^{-1}$ in Fig. 3(f). This value coincides with the skyrmion lattice distance $d = (\sqrt{3}/2)a_{sk} \sim 70$ nm, where $a_{sk} \sim 80$ nm is the distance between two skyrmions measured from the Lorentz image in Fig. 3(c). We note that the reciprocal space maps obtained by the Fourier transform of Figs. 3(a)–(d) can also give information on the orientation of the SkL, similar to Figs. 3(e)–(h). However, the diffraction method can independently identify the period of the SkL and the intensity of the diffraction patterns relates to the magnitude of the magnetization[28, 29, 36]. Meanwhile, it is difficult to quantitatively analyze the reciprocal maps obtained by the Fourier transformation because the intensity of a Fresnel image is nonlinear with respect to the magnetization[25, 26].

When the two SkL domains encounter each other during the application of the magnetic field, the SkL domains rotated to account for the difference in their orientations of their lattices, as shown in Fig. 3(d). No SkL domain boundary was created between these domains with different orientations. The SmAED pattern obtained from the region D is shown in Fig. 3(h). The angle of diffraction spots from the horizontal line is ~ 4.0°, and this angle is close to the angle of the left SkL domain B (2.0°). This SkL domain was influenced by the SkL domain B because the SkL domain B was populated in a larger area. These observations show that if the SkL domains are formed in a wide area, the SkL domains will rotate without forming a SkL domain boundary when the SkL domains with different domains collide. In addition, we observed that at the bottom of the SkL domain in Fig. 3(d), a SkL domain boundary was created along a grain boundary. In the vicinity of the grain boundary, skyrmions are aligned along the grain boundary, similar to Fig. 1. Thus, this observation also supports the theory that a SkL domain boundary can be formed by the effects of grain boundaries.

We discuss the structural characteristics of the SkL domain boundary observed in Fig 1(e). The rotation angle between the two SkL domains in Fig. 1(e) is $2\theta = 88.9° \approx 90°$. In bicrystals of atomic structures with hexagonal symmetry, several types of boundaries such as $\Sigma 7\{1\bar{1}02\}$ and $\{10\bar{1}0\}$ have been reported. However, the observed SkL domain boundary structure in Fig. 1(d) is not seen in bicrystals of atomic structures[37, 38]. The bicrystals of atoms have periodic units formed at the grain boundaries[37]. Unlike bicrystals, the SkL domain boundary has irregular arrangements of skyrmions and no periodic structure is observed at the SkL domain boundary. Thus, this structure is different from that expected from the analogy to atoms in the crystal lattice, and it is considered to be a unique structural boundary in SkL domains. Furthermore, a recent study of skyrmions in $Cu_2OSeO_3$ has shown that the SkL domain boundary has no preferred tilt angle between two lattices[13]. Therefore, this study suggests that tilt angles in the SkL domain boundary can be determined to lower the total energy by other factors such as grain boundaries in a chiral magnet. The rotation angle ≈ 90° corresponds to the angle formed by the two grain boundaries [red lines in Fig. 1(c)], thus showing that the grain boundaries result in the formation of the SkL domain boundary.

In addition, we compare the domain boundary of fluxons induced by defects[22] with the observed SkL domain boundary. The domain boundaries in fluxons are formed by changing the positions of several fluxons and the domain boundary is not clear because fluxons are not regularly distributed in each domain. Conversely, the SkL domain boundary is formed by a few skyrmions [Fig. 1(e)] and skyrmions are arranged regularly in other areas [Fig. 1(d)]. Therefore, this property of the SkL domain boundary is specific to the particle-like nature of skyrmions, which are induced by the DM interaction.

In conclusion, we observed the formation process of SkL domain boundaries using Lorentz microscopy. We showed that SkL domain boundaries form in areas confined by grain boundaries because skyrmions are aligned along grain boundaries. The stability of the SkL domains in a confined area was confirmed by numerical simulations based on the Landau−Lifshitz−Gilbert equation. In addition, we observed that the particle-like nature of skyrmions appeared across the phase transition from the helical to skyrmion states. We also observed that SkL domains rotated under the application of an increasing magnetic field and no such boundary was created when skyrmions with different orientations collided in a wide area far from grain boundaries.

**SUPPLEMENTARY MATERIAL**

See supplementary material for the reproduction of the SkL domain boundary and temperature dependence of the magnetization.

**ACKNOWLEDGEMENTS**

We thank T. Kimura (the University of Tokyo) and Y. Togawa (Osaka Prefecture University) for fruitful discussions and I. Yamada (Osaka Prefecture University) for the high pressure synthesis. This work was partially supported by JSPS KAKENHI (Nos. 17H02924, 16H03833, and 15K13306) and grants from the Murata Foundation. M. M. also thanks supports from Waseda University Grant for Special Research Projects (Project No. 2017S-101) and JST PRESTO (Grant No. JPMJPR132A).

**Figures**

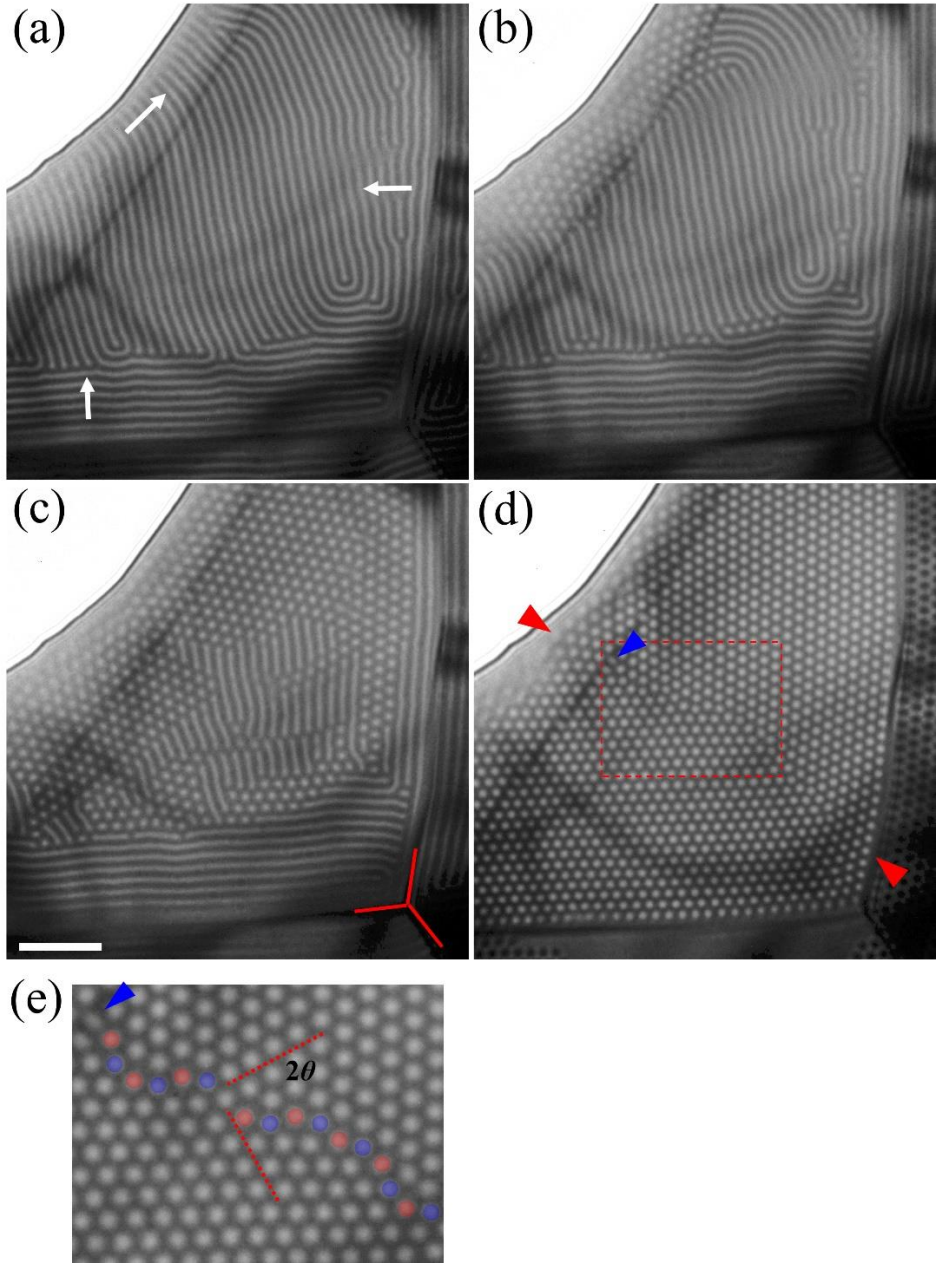

Fig. 1. Formation process of SkL domain boundaries in the area surrounded by grain boundaries at (a) 0 mT, (b) 45 mT, (c) 60 mT, and (d) 155 mT. The images are taken at 260 K. The defocus value is −6 μm (underfocus). The white arrows represent the propagation vectors of the helical domains in panel (a). The red lines in panel (c) indicate grain boundaries. The blue arrowhead represents an elongated skyrmion and the red arrowheads show an SkL domain boundary in panel (d). The SkL domain boundary comprises skyrmions with five and seven neighbors. (e) Magnified image marked by the dashed rectangle in panel (d). Skyrmions with five and seven neighbors are indicated by the red and blue spheres, respectively. The scale bar is 500 nm.

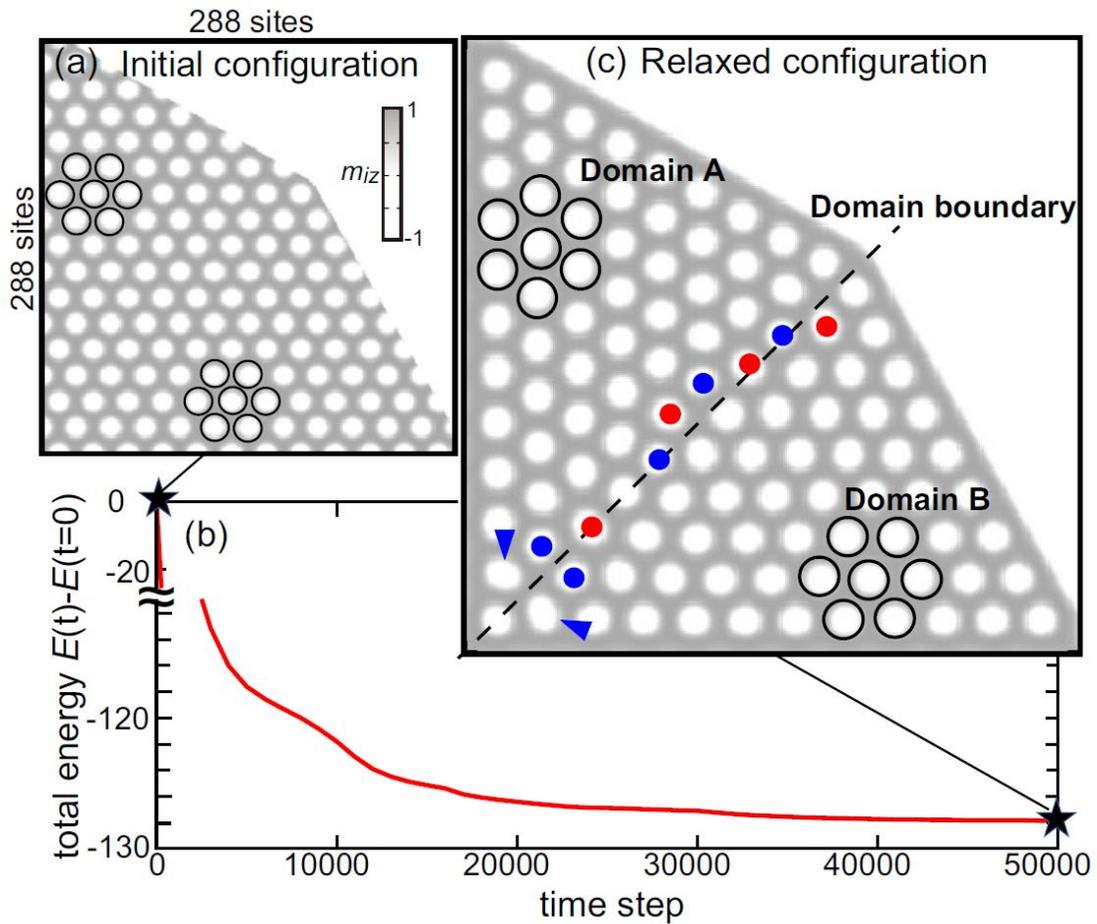

Fig. 2. (a) Single skyrmion-lattice domain confined in a system of a specific shape with an open boundary condition. (b) Simulated evolution of the total energy in the relaxation process. (c) Relaxed structure of two different skyrmion lattice domains separated by a domain boundary. Skyrmions with five and seven neighbors are indicated by blue and red spheres, respectively, while elongated skyrmions are indicated by blue arrowheads.

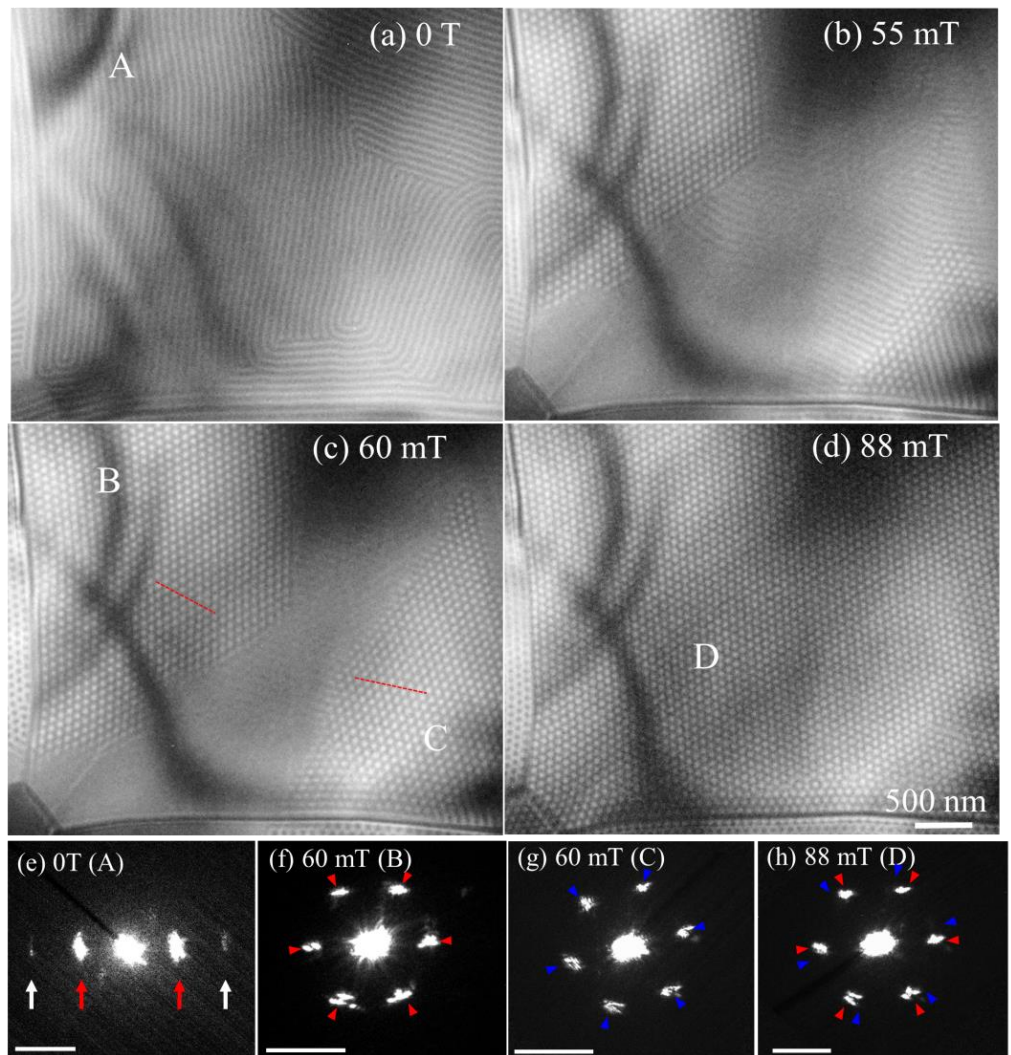

Fig. 3. (a)–(d) Overfocused Fresnel images with increasing magnetic field at 260 K. The defocus value is +5 μm. The letters A–D indicate the positions where SmAED patterns were obtained. The dashed lines represent the orientations of the SkL domains in panel (c). SmAED patterns at (e) 0 T from the region A, (f) 60 mT from region B, (g) 60 mT from region C, and (h) 88 mT from region D. Each pattern was taken from an area with a diameter of ~2 μm. The red and blue arrowheads represent the positions of the diffraction spots in panels (f) and (g), respectively. The camera length for the electron diffraction was approximately 100 m. In panels (f) and (g), a pair of diffraction spots of $3.6 \times 10^{-5}$ rad corresponds to a lattice period of $(70\ \text{nm})^{-1}$ in the reciprocal space. The scale bars are $4.0 \times 10^{-5}$ rad in the diffraction patterns.